\documentclass[prl,twocolumn,superscriptaddress,showpacs,preprintnumbers,amsmath,amssymb,nofootinbib]{revtex4}
\usepackage[dvipdfmx]{graphicx}
\usepackage{dcolumn}
\usepackage{bm}
\usepackage[dvipdfmx]{color}

\newcommand{\Slash}[1]{{\ooalign{\hfil/\hfil\crcr$#1$}}}
\def\Slash#1{\not\!\!#1}

\usepackage{color}

\begin{document}

\title{Matter-antimatter coexistence method for finite density QCD \\
toward a solution of the sign problem} 

\author{Hideo~Suganuma}
\affiliation{Department of Physics, Kyoto University, \\
 Kitashirakawaoiwake, Sakyo, Kyoto 606-8502, Japan\\
\email{suganuma@scphys.kyoto-u.ac.jp}
}

\date{\today}

\begin{abstract}
Toward the lattice QCD calculation at finite density, 
we propose ``matter-antimatter coexistence method'',  
where matter and anti-matter systems are prepared 
on two parallel ${\bf R}^4$-sheets in five-dimensional Euclidean space-time.
We put a matter system $M$ with a chemical potential $\mu \in {\bf C}$ 
on a ${\bf R}^4$-sheet, 
and also put an anti-matter system $\bar M$ with $-\mu^*$ 
on the other ${\bf R}^4$-sheet shifted in the fifth direction. 
Between the gauge variables $U_\nu \equiv e^{iagA_\nu}$ in $M$ and 
$\tilde U_\nu \equiv e^{iag \tilde A_\nu}$ in $\bar M$, 
we introduce a correlation term $S_\lambda \equiv 
\sum_{x,\nu} 2\lambda  \{N_c-{\rm Re~tr} [U_\nu(x) \tilde U_\nu^\dagger(x)]\} 
\simeq \sum_x \frac{1}{2}\lambda a^2 \{A_\nu^a(x)-\tilde A_\nu^a(x)\}^2$  
with a real parameter $\lambda$. 
In one limit of $\lambda \rightarrow \infty$, 
a strong constraint $\tilde U_\nu(x)=U_\nu(x)$ is realized, 
and therefore the total fermionic determinant becomes real and non-negative, 
due to the cancellation of the phase factors in $M$ and $\bar M$, 
although this system resembles QCD with an isospin chemical potential.
In another limit of $\lambda \rightarrow 0$, this system goes to 
two separated ordinary QCD systems 
with the chemical potential of $\mu$ and $-\mu^*$. 
For a given finite-volume lattice, 
if one takes an enough large value of $\lambda$, 
$\tilde U_\nu(x) \simeq U_\nu(x)$ is realized and   
phase cancellation approximately occurs 
between two fermionic determinants in $M$ and $\bar M$, 
which suppresses the sign problem 
and is expected to make the lattice calculation possible.  
For the obtained gauge configurations of the coexistence system, 
matter-side quantities are evaluated 
through their measurement only for the matter part $M$. 
The physical quantities in finite density QCD are expected to be estimated 
by the calculations with gradually decreasing $\lambda$ and 
the extrapolation to $\lambda=0$.
We also consider more sophisticated improvement of this method 
using an irrelevant-type correlation.
\end{abstract}

\maketitle

\def\slash#1{\not\!#1}
\def\slashb#1{\not\!\!#1}
\def\slashbb#1{\not\!\!\!#1}

\section{I. Introduction}

Nowadays, quantum chromodynamics (QCD) has been established 
as the fundamental theory of strong interaction. 
Together with the success of perturbative QCD 
for high-energy processes of hadron reactions, 
the lattice QCD Monte Carlo simulation has been 
a powerful tool to analyze nonperturbative aspects of QCD, 
after the formulation of lattice QCD and its numerical success \cite{C7980}.
Indeed, lots of studies of the QCD vacuum, hadrons and the quark-gluon plasma 
have been done for both zero-temperature and finite-temperature in lattice QCD.

Finite density QCD is also important to understand the QCD diagram, 
nuclear systems and neutron stars, and it is desired to perform 
the lattice QCD analysis as the first-principle calculation of the strong interaction. 
However, there appears a serious problem called the ``sign problem'' \cite{P83,FK02} 
in the practical lattice QCD calculation at finite density.
This problem originates from the complex value including minus sign 
of the QCD action and the fermionic determinant at finite density 
even in the Euclidean metric \cite{HK84}. 
 
At finite density with the chemical potential $\mu$, 
the Euclidean QCD action $S[A, \psi, \bar \psi;\mu]$ 
is generally complex, 
\begin{eqnarray}
S[A, \psi, \bar \psi;\mu]&=&S_G[A]
+\int d^4x \{\bar \psi (\Slash D+m+\mu\gamma_4) \psi \} \cr
&\in& {\bf C},
\end{eqnarray}
with the gauge action $S_G[A] \in {\bf R}$ and 
covariant derivative $D^\nu \equiv \partial^\nu +igA^\nu$. 
(In this paper, we use hermite $\gamma$-matrices of 
$\gamma_\mu^\dagger=\gamma_\mu$ in the Euclidean metric.)
Therefore, one cannot identify the action factor as a probability density in the QCD generating functional, unlike ordinary lattice QCD calculations. 
Also, the fermionic determinant at finite density generally takes 
a complex value \cite{HK84}, 
and its phase factor is drastically changed depending on the gauge configuration 
in a large-volume lattice, 
which makes numerical analyses difficult. This is the sign problem. 

In this paper, we propose a new method of 
``matter-antimatter coexistence method'' \cite{S17} 
utilizing the charge conjugation symmetry 
for the practical lattice QCD calculation at finite density, 
aiming at a possible solution of the sign problem. 
The organization of this paper is as follows. 
In Sec.~II, we propose a new theoretical method of 
``matter-antimatter coexistence method'' in Euclidean QCD at finite density,
and show its actual procedure for the lattice calculation.
Section~III will be devoted to the summary and the conclusion.

\section{II. Matter-Antimatter Coexistence Method}

In this section, we introduce the matter-antimatter coexistence method 
for general complex chemical potential $\mu \in {\bf C}$.
In this method, we use phase cancellation of 
the fermionic determinants between 
a matter system with $\mu$ and an anti-matter system with $-\mu^*$, 
which generally holds in QCD at finite density.

\subsection{A. General property of QCD at finite density}

To begin with, we start from the general property of 
finite-density QCD \cite{HK84},
\begin{eqnarray}
S[A, \psi, \bar \psi; \mu]^*=S[A, \psi, \bar \psi; -\mu^*],
\label{eq:action}
\end{eqnarray}
for the Euclidean QCD action $S[A, \psi, \bar \psi; \mu]$
in the presence of the chemical potential $\mu \in {\bf C}$.

For instance, in continuum QCD, one finds
\begin{eqnarray}
[\bar \psi (\Slash D+m+\mu \gamma_4) \psi ]^*
=\bar \psi (\Slash D+m-\mu^*\gamma_4) \psi,
\end{eqnarray}
which leads to Eq.(\ref{eq:action}) and 
\begin{eqnarray}
{\rm Det} (\Slash D+m+\mu \gamma_4)^*
={\rm Det} (\Slash D+m-\mu^*\gamma_4).
\end{eqnarray}

Also in lattice QCD, 
the fermionic kernel $D_F$ corresponding to ${\Slash D}+m$ generally satisfies  
$D_F^\dagger=\gamma_5 D_F \gamma_5$ \cite{R12}, 
and therefore one finds 
\begin{eqnarray}
[\bar \psi (D_F+\mu \gamma_4) \psi ]^*
=\bar \psi (D_F-\mu^*\gamma_4) \psi,
\end{eqnarray}
which leads to Eq.(\ref{eq:action}) and 
\begin{eqnarray}
{\rm Det} (D_F+\mu \gamma_4)^*
={\rm Det} (D_F-\mu^*\gamma_4).
\label{eq:det}
\end{eqnarray}

Then, as an exceptional case, 
the QCD action with the pure imaginary chemical potential $\mu \in i{\bf R}$ 
is manifestly real, and hence its lattice calculation is free from the sign problem. 
However, the QCD action is generally complex at finite density.

\subsection{B. Definition and setup of matter-antimatter coexistence method}

Now, we show the definition and setup of our approach, 
the ``matter-antimatter coexistence method'' \cite{S17}. 
In this method, we consider matter and anti-matter systems 
on two parallel ${\bf R}^4$-sheets in five-dimensional Euclidean space-time.
For the matter system $M$ with a chemical potential $\mu \in {\bf C}$ 
on a ${\bf R}^4$-sheet, we also prepare the anti-matter system $\bar M$ with $-\mu^*$ 
on the other ${\bf R}^4$-sheet shifted in the fifth direction, as shown in Fig.~1.

\begin{figure}[h]
\begin{center}
\includegraphics[scale=0.72]{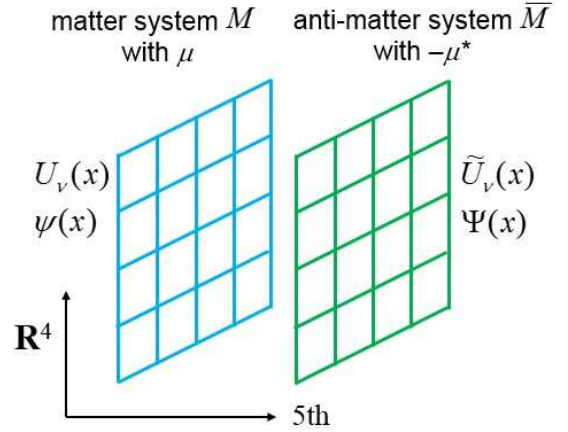}
\caption{
The matter-antimatter coexistence system in five-dimensional Euclidean space-time.
We put the matter system $M$ with $\mu$, $U_\nu(x)$ and $\psi(x)$  
on a ${\bf R}^4$-sheet, and the anti-matter system $\bar M$ 
with $-\mu^*$, $\tilde U_\nu(x)=U_\nu(x+\hat 5)$ and $\Psi(x)=\psi(x+\hat 5)$ 
on the other ${\bf R}^4$-sheet shifted in the fifth direction.
}
\end{center}
\end{figure}

We put an ordinary fermion field $\psi(x)$ with the mass $m$ 
and the gauge variable $U_\nu (x) \equiv e^{iag A_\nu(x)}$ 
at $x \in {\bf R}^4$ on the matter system $M$, 
and put the other fermion field $\Psi(x) \equiv \psi(x+\hat 5)$ 
with the same mass $m$ and the gauge variable 
$\tilde U_\nu(x) \equiv e^{iag \tilde A_\nu(x)} \equiv U_\nu (x+\hat 5)$ 
on the anti-matter system $\bar M$.
Here, $\hat 5$ denotes the fifth direction vector with an arbitrary length $a_5$, 
which is independent of four-dimensional lattice spacing $a$.

Between the gauge variables $U_\nu(x)$ in $M$ 
and $\tilde U_\nu(x)$ in $\bar M$, 
we introduce a correlation term such as
\begin{eqnarray}
S_\lambda \equiv \sum_{x,\nu} 2\lambda 
\{N_c-{\rm Re~tr} [U_\nu(x) \tilde U_\nu^\dagger(x)]\}
\label{eq:corr}
\end{eqnarray}
with a real parameter $\lambda$ ($\ge 0$) in the lattice formalism.
This correlation acts on $U_\nu(x)$ and $\tilde U_\nu(x)$ 
at the same four-dimensional coordinate $x \in {\bf R}^4$, 
and connects two different situations: 
$\tilde U_\nu(x)=U_\nu(x)$ in $\lambda \rightarrow \infty$ 
and two separated QCD systems in $\lambda \rightarrow 0$.

In fact, the total lattice action in this method is expressed as
\begin{eqnarray}
S&=&S_G[U]+\sum_x \bar \psi(D_F[U]+\mu\gamma_4)\psi \cr
&+&S_G[\tilde U]+\sum_x \bar \Psi(D_F[\tilde U]-\mu^* \gamma_4)\Psi \cr
&+&\sum_{x,\nu} 
2\lambda \{N_c-{\rm Re~tr} [U_\nu(x) \tilde U_\nu^\dagger(x)]\}
\end{eqnarray}
with the gauge action $S_G[U] \in {\bf R}$ 
and the fermionic kernel $D_F[U]$ in lattice QCD.
After integrating out the fermion fields $\psi$ and $\Psi$, 
the generating functional of this theory reads 
\begin{eqnarray}
Z&=&\int DU e^{-S_G[U]}{\rm Det} (D_F[U]+\mu\gamma_4) \cr
&& \int D{\tilde U} e^{-S_G[\tilde U]}{\rm Det} (D_F[\tilde U]-\mu^* \gamma_4) \cr
&& e^{-\sum_{x,\nu} 2\lambda \{N_c-{\rm Re~tr} [U_\nu(x) \tilde U_\nu^\dagger(x)]\}} \cr
&=&\int DU \int D{\tilde U} e^{-(S_G[U]+S_G[\tilde U])} \cr
&& {\rm Det} \{(D_F[U]+\mu\gamma_4) (D_F[\tilde U]-\mu^* \gamma_4)\} \cr
&& e^{-\sum_{x,\nu} 2\lambda 
\{N_c-{\rm Re~tr} [U_\nu(x) \tilde U_\nu^\dagger(x)]\}}.
\end{eqnarray}

\subsection{C. Four-dimensional continuum limit}

Near the four-dimensional continuum limit of $a \simeq 0$, 
this additional term becomes 
\begin{eqnarray}
S_\lambda &\simeq& 
\sum_x \frac{1}{2}\lambda a^2 \{A_\nu^a(x)-\tilde A_\nu^a(x)\}^2 \cr
&\simeq& \int d^4x~\frac{1}{2} \lambda_{\rm phys} \{A_\nu^a(x)-\tilde A_\nu^a(x)\}^2
\end{eqnarray}
with $\lambda_{\rm phys} \equiv \lambda a^{-2}$, 
and the generating functional goes to
\begin{eqnarray}
Z_{\rm cont}&=&\int DA \int D{\tilde A}  e^{-(S_G[A]+S_G[\tilde A])} \cr
&&{\rm Det} \{(\Slash{D}+m+\mu\gamma_4) (\Slash{\tilde D}+m-\mu^* \gamma_4)\} \cr
&& e^{-\int d^4x \frac{1}{2}\lambda_{\rm phys} \{A_\nu^a(x)-\tilde A_\nu^a(x)\}^2}
\end{eqnarray}
with the continuum gauge action $S_G[A] \in {\bf R}$ and 
$\tilde D^\nu \equiv \partial^\nu+i g\tilde A^\nu$. 

\subsection{D. Lattice calculation procedure}

Next, we show the actual procedure of this method 
for the lattice QCD calculation at finite density.
In the practical lattice calculation with the Monte Carlo method, 
the fermionic determinant in $Z$  
is factorized into its amplitude and phase factor as
\begin{eqnarray}
Z&=&\int DU \int D{\tilde U} e^{-(S_G[U]+S_G[\tilde U])} \cr
&& |{\rm Det} \{(D_F[U]+\mu\gamma_4) (D_F[\tilde U]-\mu^* \gamma_4)\}| \cr
&& e^{-\sum_{x,\nu} 2\lambda \{N_c-{\rm Re~tr} [U_\nu(x) \tilde U_\nu^\dagger(x)]\}} \cr
&& O_{\rm phase}[U,\tilde U],
\end{eqnarray}
and the phase factor of the total fermionic determinant  
\begin{eqnarray}
O_{\rm phase}[U,\tilde U]
\equiv 
e^{i{\rm arg}[{\rm Det} \{(D_F[U]+\mu\gamma_4) (D_F[\tilde U]-\mu^* \gamma_4)\}]}
\label{eq:phase}
\end{eqnarray}
is treated as an ``operator'' instead of a probability factor, 
while all other real non-negative factors in $Z$ can be 
treated as the probability density.

The additional term $S_\lambda$ connects the following 
two different situations as the two limits of the parameter $\lambda$:

\noindent
1) In one limit of $\lambda \rightarrow \infty$, 
a strong constraint $\tilde U_\nu(x)=U_\nu(x)$ is realized, 
and the phase factors of two fermionic determinants 
${\rm Det} (D_F[U]+\mu\gamma_4)$ and 
${\rm Det} (D_F[\tilde U]-\mu^* \gamma_4)$
are completely cancelled, owing to Eq.~(\ref{eq:det}).
Therefore, the total fermionic determinant is real and non-negative,
\begin{eqnarray}
{\rm Det} \{(D_F[U]+\mu\gamma_4) (D_F[\tilde U=U]-\mu^* \gamma_4)\} \ge 0,
\end{eqnarray} 
and the numerical calculation becomes possible without the sign problem. 
Note however that this system resembles QCD  
with an isospin chemical potential \cite{SS01}, 
which is different from finite density QCD.

\noindent
2) In another limit of $\lambda \rightarrow 0$, this system goes to 
``two separated ordinary QCD systems''  
with the chemical potential of $\mu$ and $-\mu^*$, 
although the cancellation of the phase factors cannot be expected between 
the two fermionic determinants 
${\rm Det} (D_F[U]+\mu\gamma_4)$ and 
${\rm Det} (D_F[\tilde U]-\mu^* \gamma_4)$ 
for significantly different $U_\nu(x)$ and $\tilde U_\nu(x)$, which are 
independently generated in the Monte Carlo simulation.

In fact, in other words, this approach links 
QCD with a chemical potential and 
QCD with an isospin-chemical potential. 

For a given four-dimensional finite-volume lattice, 
if one takes an enough large value of $\lambda$, 
$\tilde U_\nu(x) \simeq U_\nu(x)$ is realized,
and approximate phase cancellation occurs  
between the two fermionic determinants 
${\rm Det} (D_F[U]+\mu\gamma_4)$ and 
${\rm Det} (D_F[\tilde U]-\mu^* \gamma_4)$ in $M$ and $\bar M$.
Then, we expect a modest behavior of the phase factor 
$O_{\rm phase}[U,\tilde U]$ in Eq.(\ref{eq:phase}), 
which leads to feasibility of the numerical lattice calculation 
with suppression of the sign problem.
 
Once the lattice gauge configurations of the coexistence system are obtained 
with the most importance sampling in the Monte Carlo simulation, 
matter-side quantities can be evaluated 
through their measurement only for the matter part $M$ with $\mu$. 

By performing the lattice calculations with gradually decreasing $\lambda$ 
and their extrapolation to $\lambda=0$, 
we expect to estimate the physical quantities in finite density QCD 
with the chemical potential $\mu$.
(This procedure may resemble the chiral extrapolation, where the current quark mass 
$m$ is gradually reduced and the lattice data is extrapolated to $m=0$.)

\subsection{E. More sophisticated correlation between matter and antimatter systems}

As a problem in this method, there could appear 
an obscure effect from the additional correlation at the finite value of $\lambda$.
On this point, we here consider a possible remedy in this framework.

So far, we have demonstrated this method 
by taking the simplest correlation of $S_\lambda$ in Eq.(\ref{eq:corr}).
In this method, however, 
there is some variety on the choice of the correlation between 
$U_\nu(x)$ in $M$ and $\tilde U_\nu(x)$ in $\bar M$. 

In fact, the validity of the data extrapolation can be checked 
by various extrapolations with different type of the additional correlation. 

In particular, it is interesting to consider more sophisticated correlation such as 
\begin{eqnarray}
\bar S_\xi &\equiv& \sum_{x} 8\xi
\left( \sum_\nu \{N_c-{\rm Re~tr} [U_\nu(x) \tilde U_\nu^\dagger(x)]\} \right)^3 \cr
&\simeq& \int d^4x~\frac{1}{8} a^2 \xi [\{A_\nu^a(x)-\tilde A_\nu^a(x)\}^2]^3
\end{eqnarray}
with a dimensionless non-negative real parameter $\xi$.
At the classical level, this correlation is an irrelevant interaction 
and it gives vanishing contributions in the continuum limit $a \rightarrow 0$, 
like the Wilson term $-\frac{1}{2} ar \bar \psi D^2\psi$ \cite{R12}.

By the use of this irrelevant-type correlation, 
the effect from the additional term is expected to be reduced
in the actual lattice calculation.

\section{III. Summary and conclusion}

We have proposed the ``matter-antimatter coexistence method'' 
toward the lattice calculation of finite density QCD. 
In this method, we have prepared matter $M$ with $\mu$ 
and anti-matter $\bar M$ with $-\mu^*$  
on two parallel ${\bf R}^4$-sheets in five-dimensional Euclidean space-time, 
and have introduced a correlation term 
$S_\lambda\equiv \sum_{x,\nu} 
2\lambda \{N_c-{\rm Re~tr} [U_\nu(x) \tilde U_\nu^\dagger(x)]\} 
\simeq \sum_x \frac{1}{2}\lambda a^2\{A_\nu^a(x)-\tilde A_\nu^a(x)\}^2$ 
between the gauge variables 
$U_\nu=e^{iag A_\nu}$ in $M$ and $\tilde U_\nu=e^{iag \tilde A_\nu}$ in $\bar M$. 
In one limit of $\lambda \rightarrow \infty$, owing to $\tilde U_\nu(x)=U_\nu(x)$, 
the total fermionic determinant is real and non-negative, 
and the sign problem is absent. 
In another limit of $\lambda \rightarrow 0$, this system goes to 
two separated ordinary QCD systems 
with the chemical potential of $\mu$ and $-\mu^*$.

For a given finite-volume lattice, 
if one takes an enough large value of $\lambda$, 
$\tilde U_\nu(x) \simeq U_\nu(x)$ is realized and 
phase cancellation approximately occurs 
between two fermionic determinants in $M$ and $\bar M$, 
which is expected to suppress the sign problem 
and to make the lattice calculation possible. 
For the obtained gauge configurations of the coexistence system, 
matter-side quantities can be evaluated 
by their measurement only for the matter part $M$.
By gradually reducing $\lambda$ and the extrapolation to $\lambda=0$, 
it is expected to obtain estimation of the physical quantities 
in finite density QCD with $\mu$.

The next step is to perform the actual lattice QCD calculation 
at finite density using this method. 
It would be useful to combine this method with the other known ways 
such as the hopping parameter expansion \cite{ASSS14}, 
the complex Langevin method \cite{P83} and the reweighting technique \cite{FK02}.
For example, if the hopping parameter expansion is utilized, 
huge calculations of the fermionic determinant can be avoided, 
and a low-cost analysis with the quenched gauge configuration becomes possible, 
since the additional term $S_\lambda$ only includes gauge variables.
In addition to the actual lattice calculation, 
the effect from the additional term is to be investigated carefully. 

Efficiency of this method would strongly depend on 
the system parameters, such as the space-time volume $V$, 
the quark mass $m$, the temperature $T$ and the chemical potential $\mu$. 
For instance, near the chiral limit $m=0$ in large $V$, 
the fermionic determinant tends to possess quasi-zero-eigenvalues, 
which permits a drastic change of the phase in the fermionic determinant, 
although the zero fermionic-determinant case would give 
no significant contribution in the QCD generating functional.
In any case, this method is expected to enlarge calculable area 
of the QCD phase diagram on $(T, \mu, m, V)$.

\section*{Acknowledgements}

The author is supported in part by the Grant for Scientific Research 
[(C)15K05076] from the Ministry of Education, Science and Technology of Japan.

\end{document}